\documentclass[twocolumn,superscriptaddress,preprintnumbers,amsmath,10pt,aps,prl,nobibnotes,longbibliography]{revtex4-1} 

\usepackage{algorithm, algorithmic}

\usepackage{amsmath}
\usepackage{amsfonts}
\usepackage{bm}
\usepackage{color}
\usepackage{verbatim}
\usepackage{graphicx}
\usepackage{lipsum}
\usepackage[utf8]{inputenc}
\usepackage{times}
\usepackage{graphicx}
\usepackage{bm}
\usepackage{amssymb,multirow}
\usepackage[dvipsnames]{xcolor}
\usepackage{soul}
\usepackage[normalem]{ulem} 
\usepackage[mathlines]{lineno}
\usepackage{bbm}
\usepackage{siunitx}
\usepackage{array}
\usepackage{upgreek}
\usepackage{hyperref}
\hypersetup{
    colorlinks=true,
    linkcolor=black,
    citecolor=black,
    urlcolor=black
    }
\usepackage{gensymb} 
\usepackage{graphicx}

\def\Veca{\mathbf{a}}
\def\Vecb{\mathbf{b}}
\def\Vecc{\mathbf{c}}

\def\Vecf{\mathbf{f}}

\def\Vecs{\mathbf{s}}

\def\Vecv{\mathbf{v}}

\def\VecA{\mathbf{A}}
\def\VecB{\mathbf{B}}
\def\VecC{\mathbf{C}}

\def\VecT{\mathbf{T}}

\begin{document}

\title{Miniaturised transmissive multi-plane light converters via laser-written\\ geometric phase holograms cascaded in glass}

\author{Un\.e~G.~B\=utait\.e}
\email{u.butaite@exeter.ac.uk}
\affiliation{Physics and Astronomy, University of Exeter, Exeter, EX4 4QL. UK.}
\author{Martynas Beresna}
\affiliation{Optoelectronics Research Centre, University of Southampton, Southampton, SO17 1BJ. UK.}
\author{David~B.~Phillips}
\email{d.phillips@exeter.ac.uk}
\affiliation{Physics and Astronomy, University of Exeter, Exeter, EX4 4QL. UK.}

\begin{abstract}
Multi-plane light converters (MPLCs) are an emerging beam shaping technology capable of deterministically mapping a basis of input spatial light modes to a different basis of output modes. The ability to perform such multi-modal spatial reformatting operations has many future applications in both classical and quantum photonics, spanning from optical communications to photonic computing and advanced imaging. In this work we fabricate miniaturised transmissive MPLCs fully-encapsulated within a fused silica glass chip using single-step 3D direct laser writing. Our approach relies on the formation of femto-second laser induced birefringent nanogratings with a spatially controllable slow-axis orientation. Multiple layers of these nanogratings are laser-written throughout the volume of the glass to create a sequence of axially separated geometric phase holograms which imprint controllable phase patterns onto circularly polarised read-out light propagating through them. We construct and test a range of proof-of-concept laser-written MPLCs operating in the visible (${\lambda=633\,\text{nm}}$). These miniature beam multiplexers are formed from up to 5 separate phase masks of width $\sim 260$\,$\upmu$m, cascaded along a total length of $\sim$\,$2.7$\,mm, thus occupying a compact volume of $\sim$\,$0.15$\,mm$^3$. We first demonstrate Hermite-Gaussian (HG) mode sorters capable of diverting the energy carried by up to 28 overlapping HG modes into spatially separated output channels. We next create a 7-mode orthogonal speckle sorter, highlighting the universal nature of the spatial transformations it is possible to encode. Finally, we show analogue optical matrix multiplications achieved by passively scattering light through these 3D structured glass elements. Our work begins to merge the concepts of free-space optics with 3D integrated photonics in glass and plots a path towards the rapid prototyping of robust monolithic MPLC technology. 
\end{abstract}

\maketitle

The spatio-temporal amplitude, phase and polarisation texture of light carries an enormous number of independent degrees-of-freedom -- representing an exceptionally high information capacity. Tapping into this resource by gaining full control over all of light's dimensions is a long standing challenge in photonics~\cite{rubinsztein2016roadmap,cristiani2022roadmap,shen2023roadmap}. In particular, the ability to individually address specific degrees-of-freedom within optical fields is crucial to unlocking the full potential of photonic technologies~\cite{mounaix2020time}. While techniques to manipulate optical fields based on their spectral or polarisation properties are relatively mature, fine control over the spatial dimensions within light fields currently lags behind. A key operation is that of a {\it spatial mode sorter}: an optical system capable of passively decomposing an incident light field into a basis of transverse spatial light modes, and redirecting the energy carried by each mode to separate locations at the output~\cite{miller2015sorting}.

\begin{figure*}[ht]
\centering
\includegraphics[width=1\linewidth]{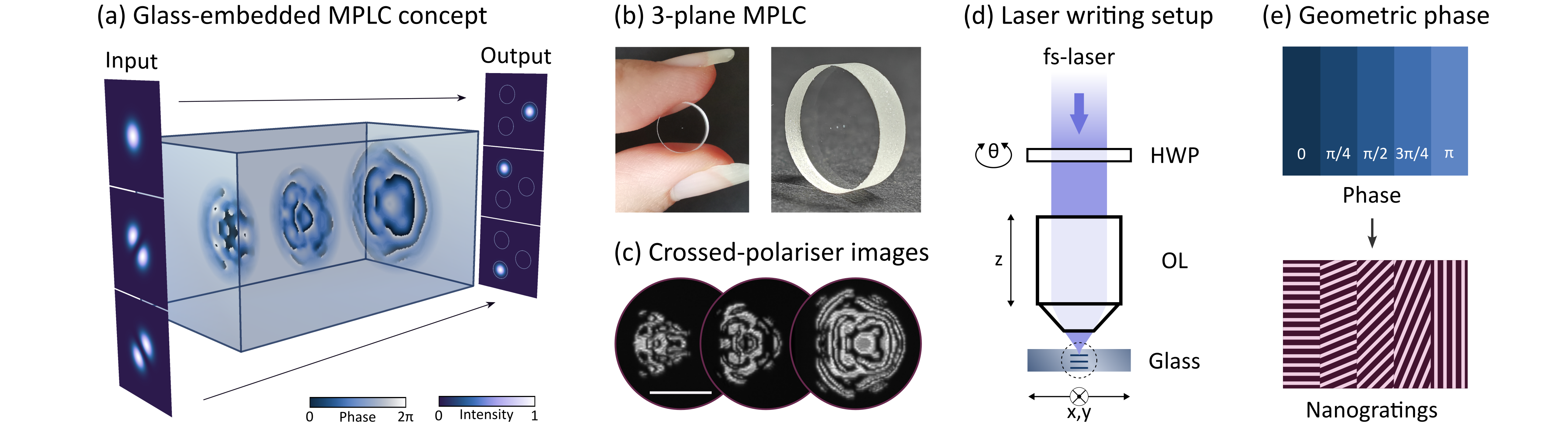}
    \caption{{\bf Concept}. (a) A schematic of a monolithic $3$-plane MPLC laser-written inside a glass chip, shown here sorting $3$ HG transverse spatial modes into focussed spots in different locations. In practice, the incident and output spatial modes are centred on a common axis -- they are shown displaced vertically here for clarity. (b) Photographs of a laser-written $3$-plane MPLC fabricated inside a $10\,\text{mm}$ diameter disk of fused silica. The separation between phase masks is $1.35$\,mm. (c) Optical microscope images of each geometric phase hologram between crossed polarisers. The scale bar is $100$\,$\upmu$m. Regions that are phase shifted by $\pi$ appear with the same intensity in this view. (d) A schematic of the custom-built direct laser writing system. The polarisation direction ($\theta$) of the linearly polarised fs-beam is controlled by a half-wave plate (HWP) held in a motorized rotation mount, prior to the beam being focussed into the glass substrate by an objective lens (OL). The glass is translated relative to the writing beam using 3D translation stages whose 3D motion (in $x$, $y$, and $z$) is synchronised with the laser emission and the HWP orientation. (e) The desired phase of each hologram pixel is mapped to the writing orientation of the laser written nanogratings, which corresponds to the local slow-axis of the geometric phase holograms.}
    \label{fig:concept}
\end{figure*}

Multi-plane light conversion is a highly versatile emerging technology that can tackle this problem~\cite{morizur2010programmable,labroille2014efficient,wang2018dynamic,fontaine2019laguerre,brandt2020high,fontaine2023wafer}. A multi-plane light converter (MPLC) -- more recently coined as a linear diffractive neural network~\cite{lin2018all,bearne2026diffractive} -- consists of a cascade of diffractive phase masks that are separated by free-space, as shown in Fig.~\ref{fig:concept}(a). The spatially-varying phase profiles of these phase masks are crafted through the process of inverse design~\cite{hashimoto2005optical} to restructure incident light as required. The optical field flowing through an MPLC is imprinted with the phase pattern of each plane in turn, while diffraction through the space between the planes serves to exchange energy laterally. In this way, input optical fields can be efficiently transformed into the desired output fields as they flow through the device. Crucially, while interaction with single-plane light shaping devices such as spatial light modulators (SLMs) or metasurfaces can typically only efficiently transform a {\it single} input spatial mode at a time~\cite{gibson2004free,vcivzmar2012exploiting,mazilu2017modal}, MPLCs can efficiently process {\it multiple} orthogonal spatial modes in parallel, thus -- in principle -- achieving near arbitrary spatial reformatting operations.

The flexibility of multi-plane light conversion has inspired rapidly growing interest in recent years. Notably, highly efficient sorting of up to 1000 Hermite-Gaussian (HG) or Laguerre-Gaussian (LG) modes has been demonstrated~\cite{fontaine2019laguerre,fontaine2021hermite}. MPLCs have also been applied to sort a range of other modal bases, including Bessel modes, Zernike modes and `random' modes formed from orthogonal sets of speckle patterns~\cite{kupianskyi2023high}. Beyond mode sorting, more general spatial transformations, basis rotations and arbitrary linear optical circuits are also made possible with this technology~\cite{brandt2020high,goel2024inverse}, opening up a wide array of potential future applications. Examples include space division multiplexing in optical communications~\cite{richardson2013space,fontaine2022photonic,rademacher20233}; quantum-optimal far-field super-resolution imaging and exoplanet detection~\cite{tsang2016quantum,rouviere2024ultra,deshler2025experimental,lvovsky2026passive}; reversing the spatial scrambling effects of multimode optical fibres~\cite{butaite2022build,kupianskyi2024all,yu2025all} and opaque scattering media~\cite{kang2023tracing,levin2025understanding}; realisation of passive optical matrix multiplication for low-energy computing architectures and optical neural networks~\cite{athale1982optical,wang2022optical,zhou2022photonic,momeni2025training}; and unitary operators for quantum computing paradigms~\cite{brandt2020high,lib2022processing,lib2024resource,goel2026quantum}.

Given the intricacy of 3D MPLCs, a major challenge is their accurate physical construction. Various methods are under development. So far, fully reconfigurable MPLCs have been realised using multiple reflections on SLMs~\cite{wang2018dynamic,fontaine2019laguerre,dinc2024multicasting,a2025self}. Fixed MPLCs have been fabricated via multi-step lithographically etched diffraction gratings~\cite{wang2020azimuthal,mounaix2020time,fontaine2023wafer} or metasurfaces~\cite{oh2022adjoint,soma2025complete}. Transmissive MPLCs have been fabricated using two photon polymerisation (2PP) 3D printing~\cite{porte2021direct,zhang2025demultiplexing,roberts20233d}. Laser writing directly within glass has also been used to engineer binary aperiodic volume optics that offer similar functionality to MPLCs~\cite{gerke2010aperiodic,barre2022inverse,barre2023direct}. Yet a straightforward approach to construct efficient high-fidelity MPLCs capable of delivering high-dimensional optical transformations in a compact form factor is still lacking.

In this work we fabricate miniaturised transmissive MPLCs within a single glass chip by laser-writing cascaded sets of geometric phase holograms. This strategy enables the creation of monolithic millimetre-scale MPLCs via a rapid-turnaround single-step process. The phase masks are automatically co-registered thus avoiding post-fabrication plane-to-plane alignment steps. We previously presented proof-of-principle experiments introducing this concept~\cite{Photon24,butaite2026miniaturised}. Here we substantially scale up the complexity and performance of this approach -- experimentally demonstrating laser-written MPLCs operating at visible wavelengths, consisting of up to $5$ phase modulation planes distributed throughout a compact glass volume of $\sim\,0.15\,\text{mm}^3$.

To showcase the broad potential of this technology, we create miniaturised HG-mode sorters processing up to 28 modes simultaneously, orthogonal speckle mode sorters, and MPLCs capable of performing arbitrary all-optical complex matrix multiplications. Complementing our work in the visible, we also highlight a recent preprint by Korichi et al.\ demonstrating a similar laser-written MPLC scheme for vector beam transformations in the infra-red~\cite{korichi2026volumetric}. We discuss the fabrication challenges and future improvements of this highly versatile glass-integrated optical technology.

\noindent{\bf MPLCs via laser-written geometric phase holograms}\\
Figure~\ref{fig:concept}(a) illustrates our concept, using a simple 3-mode HG sorter as an example. We design this mode sorter for a readout wavelength of $\lambda = 633$\,nm. The MPLC consists of $3$ phase masks, each of width $220\,\upmu$m, and separated by $1.35\,\text{mm}$. The output plane, where light from each HG mode is focussed to a distinct point, is situated $5\,\text{mm}$ after the last phase mask. Inverse design of the MPLC is carried out by optimising a digital model of the optical system using the wavefront matching method~\cite{hashimoto2005optical,fontaine2019laguerre} (see Methods). Once we have the MPLC design, we laser write the phase masks directly into a single glass chip. Figure~\ref{fig:concept}(b-c) shows photographs and crossed-polariser microscope images of the miniature laser-fabricated 3-plane MPLC, which occupies a volume of $\sim\,0.1\,\text{mm}^3$. 

\begin{figure*}[t]
    \includegraphics[width=1\linewidth]{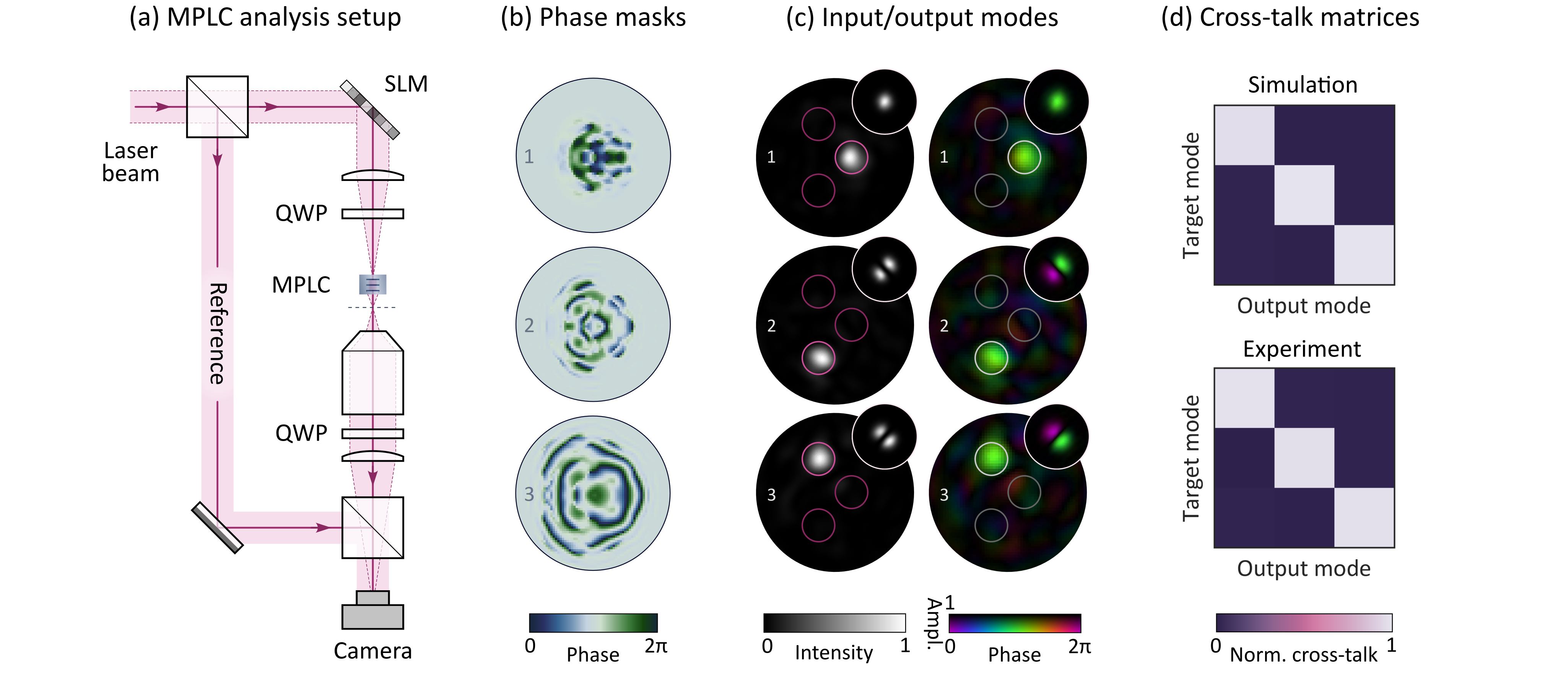}
    \caption{{\bf Characterisation of a laser-written $\bm{3}$-plane $\bm{3}$-mode HG sorter}.  (a) A schematic of the MPLC characterisation system. A liquid crystal SLM shapes the light incident on the MPLC, and  off-axis digital holography is used to measure the transmitted output field. QWP: quarter-wave plate. Supplementary Fig.~1 gives a detailed diagram of this optical setup. (b) The $3$ inverse-designed MPLC phase holograms. MPLC dimensions are: $220$\,$\upmu$m\,$\times$\,$220$\,$\upmu$m\,$\times$\,$2.7$\,mm. 
    (c) Experimentally measured input (shown in insets) and output modes when the MPLC is illuminated with 3 HG modes in turn: HG$_{\text{0,0}}$, HG$_{\text{1,0}}$, HG$_{\text{0,1}}$. Intensity is shown on the left and optical fields on the right column. The location of the output channels is marked by circles, with the target output channel highlighted. (d) Simulated and experimentally measured normalised cross-talk matrices (see Methods). Supplementary Fig.~2 shows these matrices on a log scale. The mean intermodal cross-talk of the simulated [experimental] MPLC is -20.5\,dB [-18.9\,dB].}
    \label{fig:3modeHG}
\end{figure*}

Figure~\ref{fig:concept}(d) shows a schematic of our custom-built fs-laser-writing platform. Each phase mask is fabricated by generating fs-laser induced form-birefringent nanogratings within fused silica glass~\cite{shimotsuma2003self,bhardwaj2006optically,bellouard2008scanning,richter2012nanogratings}.  We first optimise the laser writing parameters to enable the creation of a zero-order half-wave plate. The slow-axis orientation of this half-wave plate is parallel to the orientation of the nanogratings and perpendicular to the orientation of the linearly polarised writing beam~\cite{beresna2014ultrafast}. By rotating the polarisation of the writing beam as a function of lateral position in the glass substrate, we control the local orientation of the laser-written nanogratings, thus creating half-wave plates with spatially-varying slow-axis orientation which act as geometric phase holograms~\cite{beresna2011radially,drevinskas2017high}. When a circularly polarised readout beam is transmitted through these holograms, its global polarisation handedness is flipped. Accompanying this, the light beam is also imprinted with a spatially-varying geometric phase pattern (also known as a Pancharatnam-Berry phase~\cite{pancharatnam1956generalized,berry1984quantal,anandan1992geometric}), dictated by the local slow-axis orientation of the half-wave plate nanogratings, as illustrated in Fig.~\ref{fig:concept}(e) (see Methods for more detail).

We experimentally measure the transmissivity $t$ of a single phase mask, encoded with a linear phase grating and written with a continuous phase distribution, to be ${t\sim0.84}$. When using our current laser writing platform to create the more elaborate phase patterns intrinsic to MPLCs, it is necessary to round the continuous phase distribution of our designs to a discrete set of phase levels, in order to limit fabrication time. We find that using $6$ phase levels reduces the transmissivity of each phase mask to ${0.61\le t\le0.68}$ (see Methods). We emphasise that there are no fundamental bit depth constraints in our approach, so we expect high-efficiency laser-written MPLCs in the visible to be readily attainable in the future.\\

\noindent{\bf High-dimensional Hermite-Gaussian mode sorting}\\
We now test the $3$-plane $3$-mode HG sorter depicted in Fig.~\ref{fig:concept}(a-c). Figure~\ref{fig:3modeHG}(a) shows a schematic of the interferometric optical setup used to characterise the optical response of our laser-written MPLCs (see Methods for more details). In brief, an aberration corrected liquid crystal SLM shapes both the amplitude and phase of the light incident on the MPLC to generate the required input transverse spatial modes~\cite{davis1999encoding}. We employ off-axis digital holography with a coherent reference beam to reconstruct the transmitted optical field. Phase drift tracking ensures we can accurately measure the relative global phase of output fields generated by different input modes -- key for applications in which output beams will be interferometrically combined, such as the all-optical matrix multiplication examples shown later.

\begin{figure*}[ht]
    \includegraphics[width=\textwidth]{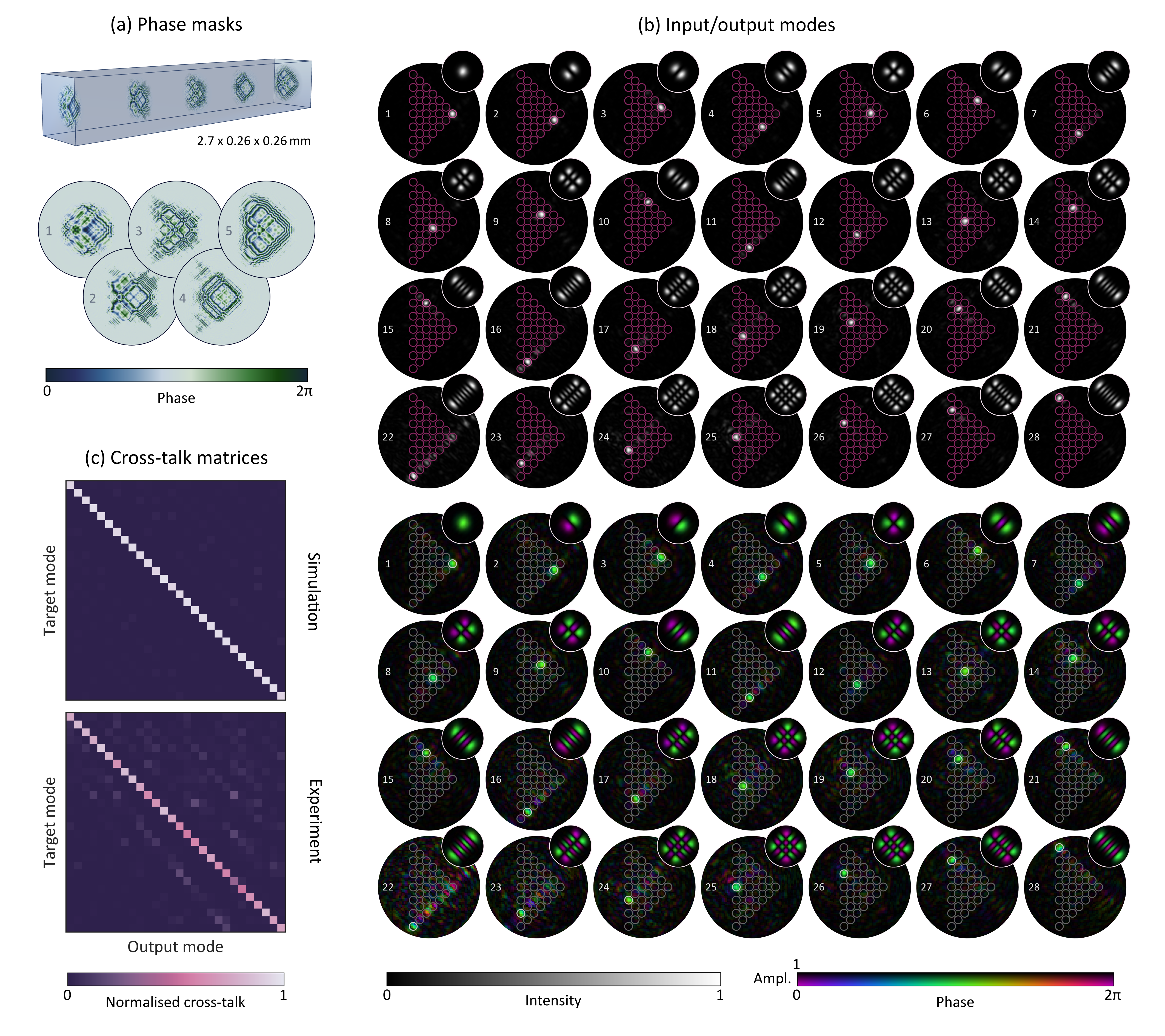}
    \caption{{\bf A $\bm{5}$-plane $\bm{28}$-mode HG sorter}. 
    (a) Upper panel: A schematic showing the cascade of $5$ laser-written holograms within the glass chip. MPLC dimensions: $260$\,$\upmu$m\,$\times$\,$260$\,$\upmu$m\,$\times$\,$2.7$\,mm.
    Lower panel: the $5$ inverse-designed phase holograms. (b) Experimentally measured input (shown in insets) and output modes when the MPLC is illuminated with 28 HG modes in turn, encompassing all HG$_{\text{m,n}}$ modes with indices satisfying $m+n\le6$. Intensity is shown in the upper four rows, and optical field in the lower four. The location of the output channels is marked by circles, with the target output channel highlighted. (c) Simulated and experimentally measured normalised cross-talk matrices. Supplementary Fig.~2 shows these matrices on a log scale. Here the mean intermodal cross-talk of the simulated [experimental] MPLC is -27.4\,dB [-18.0\,dB].}
    \label{fig:28modeHG}
\end{figure*}

Figure~\ref{fig:3modeHG}(b) shows the phase mask designs. They inherit their horizontal mirror symmetry from the symmetry of the HG modes themselves. Figure~\ref{fig:3modeHG}(c) shows the measured intensity patterns and optical field maps of the input spatial modes and the corresponding MPLC outputs. As per our design, for each input mode, we observe light predominantly focussed into the target output channel. To quantitatively analyse the sorting performance of this MPLC, Fig.~\ref{fig:3modeHG}(d) shows a comparison between simulated and experimentally measured cross-talk matrices (see Methods for cross-talk calculation).

\begin{figure*}[ht]
    \includegraphics[width=\textwidth]{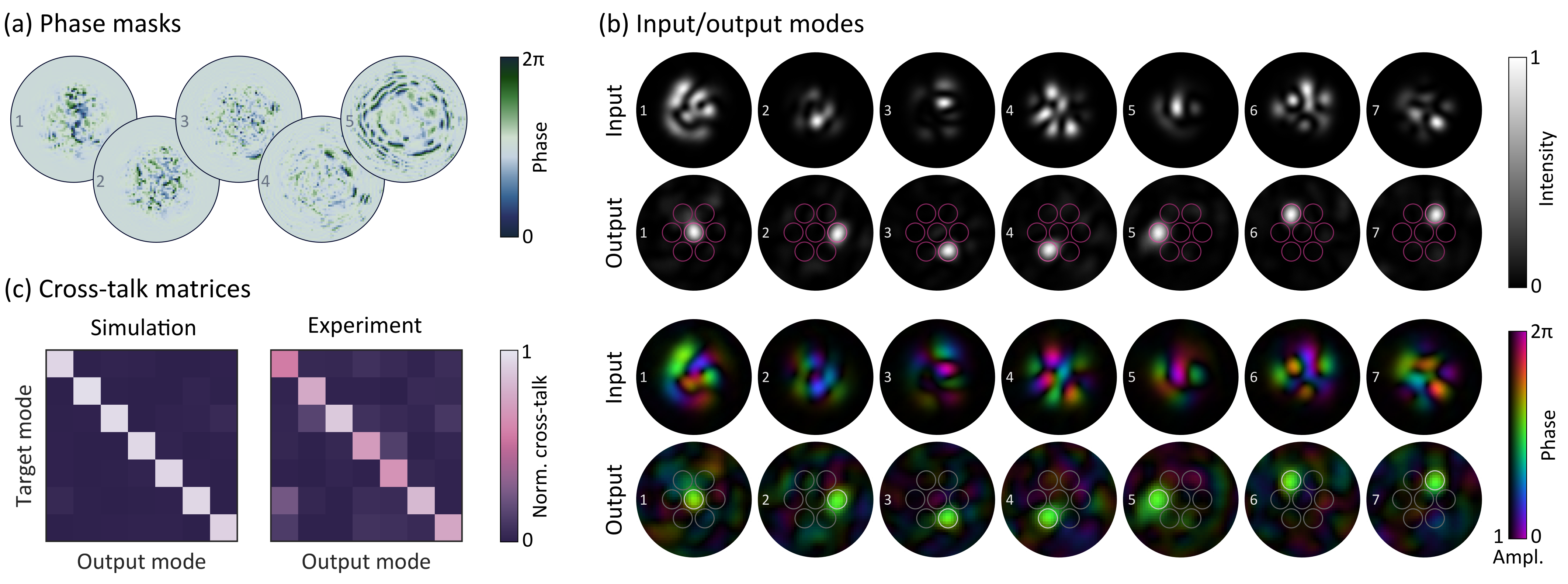}
    \caption{{\bf Arbitrary mode sorting: a $\bm{5}$-plane $\bm{7}$-mode speckle sorter}. (a) The $5$ inverse-designed phase holograms for speckle mode sorting. MPLC dimensions: $220$\,$\upmu$m\,$\times$\,$220$\,$\upmu$m\,$\times$\,$2.7$\,mm. (b) Upper two rows: experimentally measured input and output intensity patterns when the MPLC is illuminated with 7 orthogonal speckle patterns in turn. Lower two rows: experimentally measured input and output optical fields. 
    (c) Simulated and experimentally measured normalised cross-talk matrices. Supplementary Fig.~2 shows these matrices on a log scale.   Here the mean intermodal cross-talk of the simulated [experimental] MPLC is -20.2\,dB [-12.4\,dB].}
    \label{fig:speckle}
\end{figure*}

High-dimensional HG mode sorters are finding a growing number of applications, for example as spatial multiplexers for space-division multiplexing through graded-index multimode fibres~\cite{rademacher20233}, or 1D to 2D field transformers for spatio-temporal beam shaping~\cite{mounaix2020time,komonen2025programmable}. Given this, we next scale up the mode capacity and complexity of the MPLC transformation, and fabricate a proof-of-principle $5$-plane $28$-mode HG sorter with a volume of $\sim\,0.15\,\text{mm}^3$. Figure~\ref{fig:28modeHG} shows the mask designs and mode sorting performance. A higher mode capacity MPLC generally results in phase mask designs of increased complexity, which present additional challenges to fabricate accurately. Nonetheless, we observe good agreement between our simulations and the experimentally realised prototype 28-mode HG sorter. Unwanted channel cross-talk more strongly affects higher-order HG modes -- here the HG$_{6,0}$ mode proves most difficult to accurately control. We observe that sorting efficiency also decreases for higher order modes (See Supplementary Fig.~3). The Discussion section describes laser writing challenges in more detail and presents avenues for future improvement.\\

\noindent{\bf Speckle mode sorting}\\
We now turn our attention to mode sorting in an arbitrary basis. While it has been shown that it is possible to sort a large number of HG modes using a relatively low number of phase planes~\cite{fontaine2019laguerre,fontaine2021hermite}, this is a special case, and mode sorting in other bases is generally more challenging: the number of required planes scales roughly linearly with the number of modes, unless the efficiency of mode control is deliberately reduced~\cite{kupianskyi2023high}. With this typical scaling in mind, we design and fabricate a $5$-plane $7$-mode orthogonal speckle sorter. Methods gives details of how orthogonal speckle patterns are generated. Speckle mode sorting has applications, for example, in seeing through complex media by reversing complex spatial light scattering effects~\cite{kupianskyi2024all}. 

Figure~\ref{fig:speckle}(a) shows the phase mask designs, where we see that their symmetry is completely broken by the random nature of the input modes. Figure~\ref{fig:speckle}(b-c) demonstrates the performance of the speckle mode sorter, showing that laser-written MPLCs can deliver arbitrary basis rotations. \\

\noindent{\bf Analogue optical matrix multiplication}\\
Finally, we demonstrate how laser-written MPLCs have potential for all-optical information processing. There is growing interest in exploring how optical systems can be used to perform mathematical operations, with potential future benefits over conventional electronic processors including lower energy consumption and higher levels of parallelism~\cite{caulfield2010future,zhou2022photonic,momeni2025training,wright2026physical}. Matrix multiplications have been previously achieved by encoding the matrix into a single diffractive optical element~\cite{zhao2019universal,spall2020fully,kalinin2023analog}. Extending diffractive information processing to a cascade of several phase masks offers a potentially more favourable scaling in terms of light processing efficiency, especially as the dimensionality is increased~\cite{kupianskyi2023high}.

The transmission matrix of a linear optical system, such as an MPLC, is a complex matrix $\VecT$ capturing how any incident optical field $\Veca$ is transformed into a new optical field $\Vecb$ upon propagation through the system~\cite{popoff2010measuring}: 
\begin{equation}\label{Eqn:TM}
    \Vecb = \VecT\cdot\Veca.
\end{equation}
Here $\Veca$ and $\Vecb$ are represented as column vectors storing the complex coefficients expressing input and output fields in a given basis. Thus, by inverse-designing and building an MPLC to transform a particular set of input modes to a new set of output modes, we are specifying a particular matrix $\VecT$. By illuminating this MPLC with an arbitrary input field $\Veca$, and measuring the output field $\Vecb$, we are passively performing the matrix multiplication operation shown in Eqn.~\ref{Eqn:TM}.

\begin{figure*}[ht]
    \includegraphics[width=\textwidth]{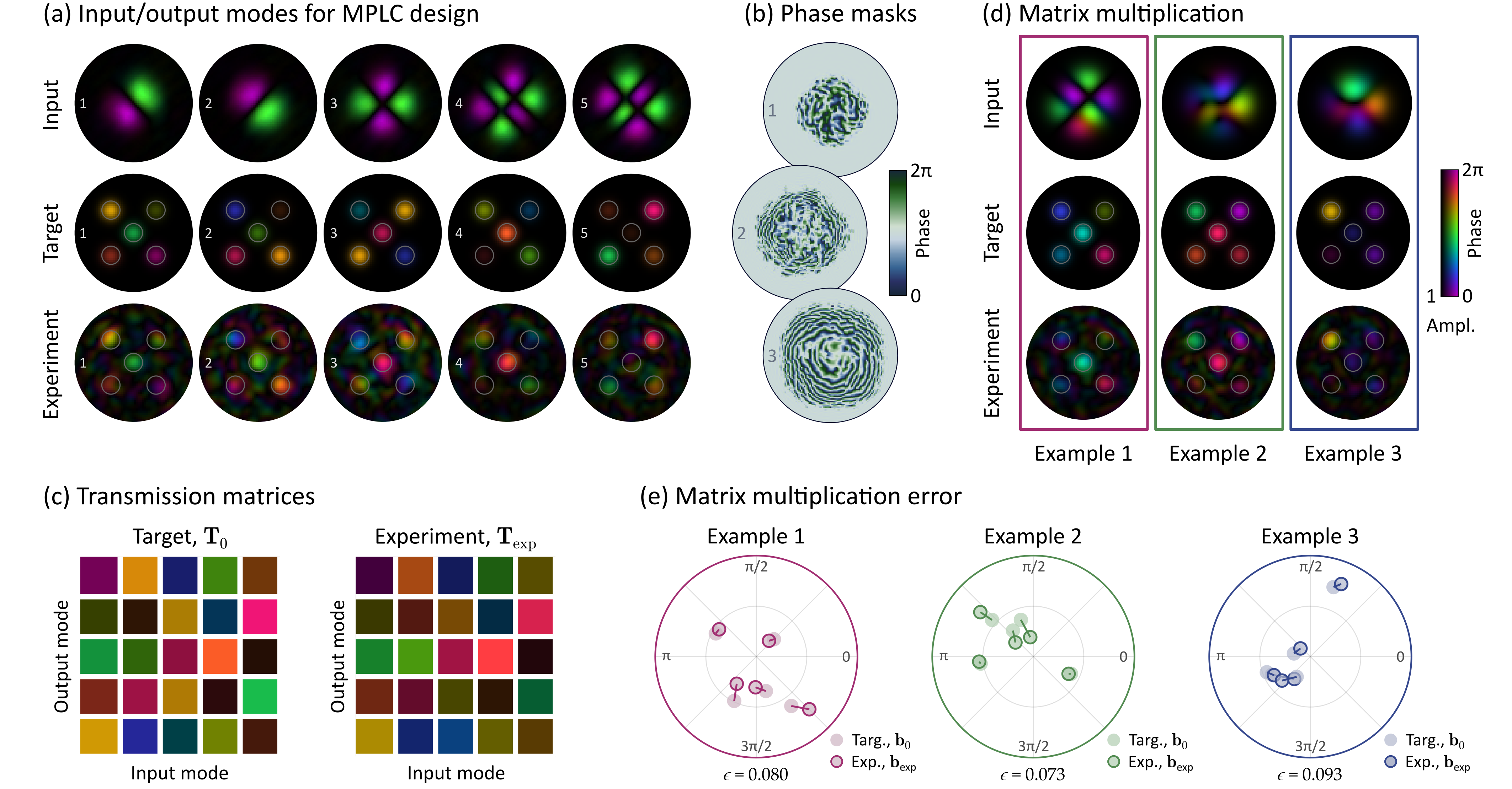}
    \caption{{\bf Analogue optical matrix multiplication using a laser written MPLC}. (a) The 5 input, target, and experimentally measured output modes. 
    (b) The $3$ inverse-designed phase holograms. MPLC dimensions: $280$\,$\upmu$m\,$\times$\,$280$\,$\upmu$m\,$\times$\,$2.7$\,mm. (c) Target and experimentally measured MPLC transmission matrices. (d) $3$ examples of optical matrix multiplications. Each column shows the input, target, and experimentally measured output fields which encode the elements of vectors $\Veca$, $\Vecb_0$, and $\Vecb_{\text{exp}}$ respectively. (e) Comparison of the complex elements of $\Vecb_0$ and $\Vecb_{\text{exp}}$ on the unit circle in the complex plane: the difference between these values is the error $\delta\Vecb$.}
    \label{fig:multiplication}
\end{figure*}

Here we test the capability of our laser-written MPLCs to emulate a target randomly defined $5\times5$ element complex unitary matrix $\VecT_{0}$, and process input light to passively achieve matrix multiplications. As shown in Fig.~\ref{fig:multiplication}(a), we define the input basis to be a set of $5$ HG modes and the output basis to be an array of $5$ focussed spots. A $3$-plane MPLC is designed to transform each input HG mode into a particular linear combination of spots, specified by coefficients held in each column of $\VecT_0$.

Figure~\ref{fig:multiplication}(b) shows the phase mask designs, which once again have broken symmetry due to the random nature of the target transformation. Once fabricated, we first measure the transmission matrix of the physically realised MPLC, $\VecT_{\text{exp}}$, in the input and output bases described above. Experimentally measured output fields are shown on the bottom row of Fig.~\ref{fig:multiplication}(a), where they can be compared with the target output fields. We observe a good match between $\VecT_{\text{exp}}$ and our target matrix $\VecT_0$, as can be seen in Fig.~\ref{fig:multiplication}(c), with the normalised overlap integral between the two matrices given by ${o = 0.88}$ (see Methods).

Figure~\ref{fig:multiplication}(d) shows $3$ examples of optical matrix multiplications carried out using this MPLC. The top row shows $3$ different randomly generated input fields, each encoding a different complex input vector. For example, an input optical field $\Vecf$ encodes the $n^{\text{th}}$ element of the input vectors $\Veca$ as the complex coefficients $a_n$ in the linear combination of HG modes: ${\Vecf = \sum_na_n\text{HG}_n}$. The middle row shows simulations of the resulting output fields in the ideal case, where the transmission matrix of the MPLC perfectly matches the target $\VecT_0$. These output fields encode the elements of ${\Vecb_0=\VecT_0\cdot\Veca}$ as the complex coefficients of the field when expressed in the spot basis. The bottom row shows the experimentally measured output fields. In reality, small differences between our target matrix $\VecT_0$ and the physically realised matrix $\VecT_{\text{exp}}$ (as seen in Fig.~\ref{fig:multiplication}(c)), result in errors in the measured output vector: ${\Vecb_{\text{exp}} = \Vecb_0 + \delta\Vecb}$. Figure~\ref{fig:multiplication}(e) shows the experimentally measured coefficients of $\Vecb_{\text{exp}}$ in comparison to those of $\Vecb_0$. In each case, we quantify the relative magnitude of this error, which is given by ${\epsilon =(\delta\Vecb^\dagger\cdot\delta\Vecb)/(\Vecb_0^\dagger\cdot\Vecb_0)}$, and find a mean error of ${\bar{\epsilon}\sim0.08}$ across the $3$ examples. We expect this error could be substantially reduced by improved fabrication, and, as in conventional electronic computation, could also be mitigated via additional redundancy or signal discretisation.\\

\noindent{\bf Discussion and conclusions}\\
We have demonstrated proof-of-concept miniaturised glass-embedded MPLCs operating in the visible and occupying volumes of ${\sim0.15\,\text{mm}^3}$. We now assess future avenues of development for this technology.

A technical limitation in our current approach is the time taken to laser-write the MPLCs. For example, the $5$-plane $28$-mode HG sorter took ${\sim10\,\text{h}}$ to fabricate. In our current laser writing platform, the fabrication time scales linearly with the number of phase levels -- a scaling imposed by the use of a motorised rotation mount for control of the linear polarisation of the writing beam. This led us to discretise the phase masks to $6$ levels thus reducing the efficiency of each phase modulation plane (see Methods). Implementing fast electro-optic polarisation control via a Pockels cell would substantially reduce the writing time, and also enable phase masks with smooth continuous variation in phase to be fabricated in a reasonable timescale~\cite{drevinskas2017high}. We expect this change would increase the transmissivity of each plane towards our measured value of ${t=0.84}$ for continuous phase modulation of a linear phase grating, thus increasing the overall efficiency of our laser written MPLCs.

Further enhancement in the light processing efficiency of the phase masks is also possible via additional optimisation of the laser writing parameters. For example, uncontrolled scattering from the nanogratings is currently the major source of loss. There are two ways forward to suppress this uncontrolled scattering: (i) by optimising the writing parameters to generate nanogratings of higher uniformity, which has been shown to enable transmissivity of up to ${t\sim0.9}$ in the visible range~\cite{drevinskas2017high,ohfuchi2017characteristic}; (ii) by switching to a recently discovered different class of birefringent glass modification known as Type X, originating from the formation of randomly distributed nanopores in silica glass~\cite{sakakura2020ultralow}. Type X modification is reported to provide ultralow scattering loss with up to ${t=0.99}$ transmission in the visible. 

In addition to efficiency improvements, there is also scope to enhance the fidelity of our laser written MPLCs. For example, one route is to build a more physically accurate digital model of the MPLC used in the inverse-design process, via the inclusion of realistic fabrication constraints. This approach was recently taken by Barré et al., who incorporated knowledge of the profile of pre-measured laser-written features into the digital model used to design high efficiency binary aperiodic volume optics~\cite{barre2023direct}. In our case, more accurately capturing the optical properties of the nanograting layers will reduce the model-reality gap in our inverse design process: our digital model currently assumes the phase masks are infinitesimally thin, while in reality the nanograting layers can be up to tens of microns thick. A physically realistic digital model will be particularly important if switching to ultra low loss Type X glass modification, as the weak birefringence of each laser-written layer means potentially tens of layers (per each MPLC plane) are required to create a half-wave plate~\cite{chavilkkadan2025purely}.

We now analyse the trade-offs encountered upon employing laser-written MPLCs at longer wavelengths. As is typical in optical system design, operation at longer wavelengths leads to larger devices: here phase mask width scales in proportion to $\lambda$ and plane separation scales as $\lambda^2$, thus MPLC volume scales as $\lambda^4$. However, the lower levels of Rayleigh scattering at longer wavelengths (scaling as $\lambda^{-4}$) can lead to more efficient light processing.

In this work we have focussed on laser writing half-wave plates that impart geometric phase patterns to a specific handedness of uniformly circularly polarised light. There is an intuitive way to extend this to polarisation independent operation: by channelling each circular polarisation component into two parallel laser-written MPLCs, with the left-handed phase masks defined as the phase conjugate of the right-handed phase masks, before recombining these polarisations at the output plane. A laser-written linear geometric phase grating will diffract light of left- and right-handed polarisation in opposite directions, so can act as the circular polarisation splitter and combiner at either end of this architecture. This approach doubles the chip's width, but only increases its depth by $2$ planes -- which is optimal in terms of MPLC efficiency. More generally, we note that by tuning the laser-writing parameters, other birefringent media can be formed, such as polarisation converters~\cite{beresna2011radially,marrucci2006optical}. Taking advantage of these features, vectorial laser-written MPLCs may be fabricated (see~\cite{korichi2026volumetric}), in analogy to those recently demonstrated using encapsulated reflective metasurfaces~\cite{soma2025complete}.

We now consider our work in the wider context of alternative MPLC fabrication techniques. Passive lithographically etched MPLCs are complicated to fabricate, requiring multi-step processes, but can benefit from extremely low levels of loss and achieve a broad spectral bandwidth in a compact package. For example, a wafer-scale $14$-plane reflective MPLC capable of highly efficient, high-fidelity sorting of $45$ HG modes across a $200$\,nm bandwidth centred on $1550$\,nm, was recently demonstrated~\cite{fontaine2023wafer}. This device occupies a volume of $\sim$$10$\,mm$^3$ and presently represents the state-of-the-art in terms of passive MPLC performance and mode capacity in a compact package.

Reflective metasurface-based MPLCs can be equally small-scale~\cite{oh2022adjoint}, but tend to have higher scattering losses and narrower spectral responses than their diffraction grating counterparts. This technology is still developing: low mode capacity mode sorters have been demonstrated so far (sorting $3-6$ modes). These solutions have the advantage of simultaneously manipulating both space and polarization degrees-of-freedom independently, and also offer more spectral selectivity~\cite{soma2025complete}.

Transmissive MPLCs have the potential to be even more compact~\cite{porte2021direct}. Such devices have recently been fabricated using 2PP 3D printing -- a highly versatile technique capable of creating free-standing 3D microstructures. State-of-the-art 2PP MPLCs for the visible have pixel pitches below $1$\,$\upmu$m and plane separations reduced to tens of microns, occupying volumes down to $\sim$$0.001$\,mm$^3$~\cite{zhang2025demultiplexing}. However, at these small scales, accurate fabrication via 2PP becomes very challenging (e.g.\ due to the precision of the polymerisation process and shrinkage distortion during development), and devices demonstrated so far exhibit relatively low fidelity and high levels of uncontrolled scattering losses. Furthermore, free-standing micro-scale components are also extremely fragile, so will require encapsulation before real-world deployment.

Our MPLC fabrication method sits somewhere between 2PP and lithographic approaches. While not yet as efficient as lithography, our scheme enables rapid prototyping by circumventing the need for complex multi-step fabrication. While not as compact as 2PP devices, our fabrication procedure is more controllable -- particularly in terms of bit depth -- leading to lower scattering losses and higher fidelity. With the anticipated future improvements sketched out here, we see no reason why compact glass-integrated MPLCs can't be scaled to tens of phase planes for high dimensional light manipulation across space, spectrum and polarisation.

Finally, fully encapsulated MPLCs are naturally compatible with other types of laser-written glass modifications, such as 3D waveguides and optical interconnects, suggesting the future possibility of flexible integration with intricate optical circuitry. In summary, we have demonstrated a new approach to create miniaturised transmissive MPLCs in the visible, which adds to the growing toolbox of techniques to deliver ever more complex photonic components at ever decreasing scales.

\section{Methods}

\noindent{\bf MPLC inverse design}\\
Inverse design of the MPLC phase masks is achieved by optimising a digital model. Here we use a custom-written wavefront matching algorithm to maximise the overlap between the target and actual output fields~\cite{hashimoto2005optical,fontaine2019laguerre}. Our model is based on scalar diffraction theory under the assumption that the polarisation is uniform across any given plane. We use the angular spectrum method to propagate the fields between the phase masks (which are assumed to be infinitesimally thin) accounting for the refractive index of the silica glass substrate. We optimise the MPLC designs for operation over a small range of wavelengths which helps to promote smooth phase mask designs that lead to higher efficiency operation when fabricated. \\

\noindent{\bf Geometric phase holograms}\\
The relationship between the desired phase and nanograting orientation can be understood via Jones calculus, where a transverse electric field vector is represented by a Jones vector describing the amplitude and phase of the two electric field components in Cartesian coordinates. If left-circularly polarised light $|L\rangle$ passes through a nanograting half-wave plate, described by a Jones matrix $H$, with its slow-axis oriented at an angle $\phi$, we have: 
\begin{equation}
\begin{split}
H|L\rangle& = \frac{1}{\sqrt{2}}\begin{bmatrix}\cos(2\phi)& \sin(2\phi)\\ \sin(2\phi) & -\cos(2\phi) \end{bmatrix} \begin{bmatrix}1\\\text{i}\end{bmatrix}\\
&= \frac{1}{\sqrt{2}}\begin{bmatrix} \text{e}^{\text{i}2\phi}\\ -\text{i}\text{e}^{\text{i}2\phi} \end{bmatrix} = |R\rangle\text{e}^{\text{i}2\phi}.
\end{split}
\end{equation}
We see that left-circularly polarised light gets converted into right-circularly polarised light $|R\rangle$ whilst simultaneously picking up a phase of $2\phi$. Meanwhile, if right-circularly polarised light passes through the same nanograting half-wave plate, it picks up the conjugate phase: $H|R\rangle = |L\rangle\text{e}^{-\text{i}2\phi}$. Given that our MPLC design model is based on scalar theory, we conjugate the phase of every second plane before writing, in order to take into account polarisation flipping between alternate MPLC planes.\\

\noindent{\bf Laser writing platform}\\
Our laser writing platform is built around a fs-pulsed laser source (Light Conversion, Pharos). In this work we use the first harmonic with a $1030$\,nm central wavelength, $300$\,fs pulse duration, and $200$\,kHz repetition rate. The laser beam is focussed through an objective lens (Mitutoyo Plan Apochromatic NIR 10X, 0.26NA) into a fused silica glass sample. Before being focussed, the linearly polarised laser beam passes through a zero-order half-wave plate (HWP) held in a computer controlled motorised rotation mount (Thorlabs PRM1/MZ8), allowing the orientation of the linear polarisation of the writing beam to be set. The laser is synchronised with 3-axis nanopositioning translation stages to control the relative motion between the fabrication laser beam and the glass substrate. Hardware control and synchronisation is accomplished via a custom-written application (Labview and AeroBasic). Lateral sample motion is achieved with an Aerotech ANT130XY direct drive 2-axis nanopositioning stage, and axial motion of the objective lens - by an Aerotech ANT130LZS direct drive single-axis stage.

During fabrication, we write from the last plane to the first, i.e., from the bottom of the glass chip to the top. To make sure that the response of each phase plane is close to that of a half-wave plate, each plane is written in two identical layers separated by $29$\,$\upmu$m. We found this to be a good distance to ensure that the newly written layer does not over-write the layer below it. The first phase plane is written at a depth of $\sim$$150$\,$\upmu$m. Each pixel in a phase hologram is either $1.6$\,$\upmu$m or $3.2$\,$\upmu$m in size and consists of a $2$x$2$ or $4$x$4$ grid of `points' separated by $0.8$\,$\upmu$m. Each point receives $200$ laser pulses, while the stages translate the sample at $0.8$\,mm/s.

Ideally, for high efficiency and fidelity, the phase masks would be written with a high bit depth - by setting the half-wave plate orientation before writing each pixel. However, because of the limited speed of our rotation mount, this process would take a very long time. As a compromise, we therefore choose to round the phase to 6 values - this decreases diffraction efficiency (see Methods section `MPLC light processing efficiency') but allows faster fabrication. During the writing, we set the half-wave plate orientation to correspond to one of the phase values and write all of the pixels with this phase. We then repeat the process for each phase value.\\

\noindent{\bf MPLC characterisation}\\
Figure~\ref{fig:3modeHG}(a) shows a schematic of the interferometric optical setup used to characterise the optical response of our laser-written MPLCs. The readout laser beam, of wavelength ${\lambda = 633\,\text{nm}}$, is expanded to fill a liquid crystal spatial light modulator (SLM) which shapes both the amplitude and phase of incident light to generate the required transverse spatial modes entering the MPLC~\cite{davis1999encoding}. It is crucial to create these input spatial modes with high quality, as the MPLC outputs are -- by definition -- highly sensitive to input beam fidelity. We therefore run an in-situ aberration correction routine~\cite{vcivzmar2010situ} and use the SLM to both generate the desired spatial modes and also suppress aberrations throughout the optical system prior to the MPLC.

To characterise our laser-written MPLCs, the SLM dynamically displays a sequence of phase holograms, cycling through the set of required input modes. For each input mode, we image the corresponding intensity pattern transmitted through the MPLC to the output plane, typically situated $5\,\text{mm}$ after the last MPLC plane. We also use off-axis digital holography with a coherent reference beam to reconstruct the transmitted optical field. We implement optical phase drift tracking between the measurement of different input modes, allowing the full phase-stabilised transmission matrix of our laser-written MPLCs to be accurately measured~\cite{a2025self}. This means we can precisely quantify the relative global phase shift between different output beams or channels -- which is key for applications in which output beams will be linearly combined, such as all-optical matrix multiplication or quantum information processing.

While accurate alignment between the MPLC phase masks is ensured by the high-accuracy translation stages used during fabrication, it is also crucial the accurately align the optical axis and magnification of the spatial modes coupled into the MPLC. To achieve this we first manually carry out a coarse alignment of the MPLC, which is mounted on a manually controlled ${x\text{-}y\text{-}z}$ translation stage. Next, fine alignment is achieved programmatically by dynamically changing the hologram displayed on the SLM to search over a small range of phase tilts, lateral shifts and range of beam waists, and choose the parameters that best minimise the output cross-talk.\\

\noindent{\bf Evaluation of MPLC cross-talk}\\
Element $C_{nm}$ of the cross-talk matrix $\VecC$ is found by calculating the normalised overlap integral between $\Vecv^{\text{exp}}_m$: the output field experimentally measured when the MPLC is illuminated with input mode $m$; and $\Vecv^{\text{targ}}_m$: the target output mode when the MPLC is illuminated with mode $n$: 
\begin{equation}
    C_{nm} = \frac{\left|\left(\Vecv^{\text{targ}}_n\right)^\dagger\cdot\Vecv^{\text{exp}}_m\right|^2}{\left[\left(\Vecv^{\text{targ}}_n\right)^\dagger\cdot\Vecv^{\text{targ}}_n\right]\left[\left(\Vecv^{\text{exp}}_m\right)^\dagger\cdot\Vecv^{\text{exp}}_m\right]}
\end{equation}
In order to consider only the cross-talk within the output modal subspace of interest, throughout the main paper we present the normalised cross-talk matrix, in which each column of $\VecC$ is independently normalised such that the absolute square of its elements sum to unity.

To calculate the mean intermodal cross-talk $\bar{c}$ we take the ratio $r$ between average off-diagonal and average diagonal values of the normalised cross-talk matrix, and convert it to dB scale using $\bar{c} = 10\log_{10}(r)$.
\\

\noindent{\bf MPLC light processing efficiency}\\
The overall mean light processing efficiency of an MPLC, $\eta$, is approximately given by
\begin{equation}\label{eqn:efficiency}
    \eta \sim \eta_{\text{design}}\times t^P,
\end{equation}
where ${0\le\eta_{\text{design}}\le1}$ is the theoretical mean light processing efficiency of the MPLC design (averaged over all modes) assuming each plane is perfectly transmissive, ${0\le t\le1}$ is the real-world transmissivity of each physically realised plane (i.e., the fraction of the intensity of incident light transmitted with the desired phase profile) and $P$ is the number of MPLC planes.

The theoretical mean light processing efficiency of the MPLC, $\eta_{\text{design}}$, is dependent upon (i) the number of modes to transform (decreasing with increasing mode count), (ii) the number of planes $P$ (increasing as $P$ is increased), and (iii) the bit depth of the final holograms used (decreasing as a function of bit depth). Here we pay particular attention to the bit depth -- specific examples are given below.

We find that the transmissivity $t$ of each plane is dependent upon (i) the laser writing parameters (e.g., pulse duration, repetition rate, number of pulses per point, substrate translation speed), (ii) the bit depth of the laser-written hologram and (iii) the complexity of the hologram (i.e., how rapidly the phase changes laterally across the hologram). To determine a baseline transmissivity, we first created phase gratings without phase discretisation by continuously rotating the polarisation as the fabrication beam was scanned across the sample. Once laser writing parameters are optimised, we fabricate a linear phase ramp with a grating period of $12.5\,\upmu\text{m}$. We measure the efficiency of light transmitted to the first diffraction order, as a fraction of incident light, to be $t\sim0.84$ (with the diffracted beam's polarisation flipped as expected). In this test we also quantify the ratio of light transmitted to the first and zero diffraction orders, finding $98.6\%$ of transmitted light sent to the first diffraction order, versus $1.4\%$ of transmitted light sent to the zero diffraction order (split into $0.2\%$ of the flipped polarisation state, and $1.2\%$ in the original polarisation state). From this we conclude the polarisation conversion extinction ratio of this laser-written geometric phase hologram to be $-18.5\,\text{dB}$.

At present our fabrication system has a slow motorized rotation mount for polarisation control, and the laser writing time scales roughly linearly with the number of phase levels. Therefore, in order to limit laser writing time in our current platform, we reduced the bit depth of the holograms by rounding the continuous phase patterns to a discrete number of phase levels. We measure the impact of this discretisation on the efficiency of the geometric phase holograms by fabricating discrete phase level linear geometric phase gratings with a grating period of $12.5\,\upmu\text{m}$ (i.e., the same period as the continuous case described above). For $4$ phase levels, we measure ${t=0.62}$, and for $8$ phase levels we measure ${t=0.67}$. As a balance between writing time and efficiency, we use $6$ phase levels for the proof-of-principle MPLCs presented in the main paper.\\ 

For each mode sorter, we experimentally measure the mean light processing efficiency $\eta$. This was achieved by measuring the total intensity of light transmitted to the $n^{\text{th}}$ output channel when the MPLC was illuminated with the $n^{\text{th}}$ input mode, normalised by the intensity of each experimentally generated input mode. Supplementary Fig.~3 shows these measurements. The mean light processing efficiency $\eta$ was found by averaging these ratios over all modes. 

From $\eta$ we can estimate a value of $t$ by rearranging Eqn.~\ref{eqn:efficiency}:
\begin{equation}\label{eqn:t}
    t=\left(\eta/\eta_{\text{design}}\right)^{\tfrac{1}{P}}.
\end{equation}
As described in more detail below, we find values of $t$ in the range $0.61\le t\le0.68$ per plane, which agrees well with our measurements on a single discretised linear phase grating described above. Based on this, we expect $t$ to increase towards $t\sim0.84$ by moving to continuous phase modulation, which would significantly boost the overall efficiency of our approach (specific estimates are given below). There are also other routes to further improve the transmissivity of each phase plane, as noted in the Discussion section of the main paper.\\

\noindent{\it $3$-plane $3$-mode HG sorter efficiency}\\
This MPLC has a design efficiency of ${\eta_{\text{design}}=0.67}$ with continuous phase profiles. The design efficiency decreases to ${\eta_{\text{design}}=0.5}$ after rounding to $6$-phase levels. We experimentally measure an overall efficiency (averaged over the $3$ modes) of ${\eta\sim0.13}$. These values give an estimate of the transmittance of each plane to be $t\sim0.64$ using Eqn.~\ref{eqn:t}.\\

\noindent{\it $5$-plane $28$-mode HG sorter efficiency}\\
This MPLC has a continuous phase design efficiency of ${\eta_{\text{design}}=0.47}$ which reduces to ${\eta_{\text{design}}=0.3}$ after rounding to $6$-phase levels. We experimentally measure an overall efficiency of ${\eta\sim0.0245}$, giving an estimate of the transmittance of each plane to be ${t\sim0.61}$. Here the increased complexity of each phase mask results in a slight reduction in transmissivity.\\

\noindent{\it $5$-plane $7$-mode speckle sorter efficiency}\\
This MPLC has a continuous phase design efficiency of ${\eta_{\text{design}}=0.27}$ which reduces to ${\eta_{\text{design}}=0.15}$ after rounding to $6$-phase levels. These values are lower than above, since HG sorters are known to be achievable with low plane counts, while there are no known equivalent solutions for random basis mode sorters. We experimentally measure an overall efficiency of ${\eta\sim0.022}$, resulting in an estimate of the transmittance of each plane of ${t\sim0.68}$.\\

We expect upgrading our laser fabrication platform to enable writing of continuous smoothly varying phase profiles will boost the overall efficiency of these MPLCs -- likely increasing $\eta_{\text{design}}$ towards the levels described above, and also increasing $t$ by reducing the scattering caused by abrupt changes in the orientation of the nanogratings from pixel to pixel. Evidently the overall loss is heavily dominated by $t$ (particularly in larger plane count MPLCs), so even relatively small increases in $t$ will substantially enhance the overall MPLC efficiency. Fortunately, it has already been shown that it is possible to further optimise the laser-writing parameters to substantially increase the transmissivity of each phase mask to $0.9\le t\le0.99$ in the visible and near infra-red~\cite{drevinskas2017high,ohfuchi2017characteristic,sakakura2020ultralow}. There is also scope to improve the MPLC design itself by further tweaking design constraints such as the distance between the planes and the level of smoothing applied in the optimisation process -- e.g.\ see the work of Fontaine et al.~\cite{fontaine2019laguerre} who have achieved designs for extremely high efficiency high-dimensional HG mode sorting MPLCs.\\

\noindent{\bf Generation of orthogonal speckle modes}\\
To create an orthogonal basis of $n$ speckle patterns, we follow the method described in ~\cite{kupianskyi2023high}. In brief, we first generate a random $n\times n$ unitary complex matrix and use the complex coefficients along each column of this matrix as the weights in a linear sum of $n$ HG beams. This results in a set of $n$ mutually orthogonal speckle patterns, as shown in Fig.~\ref{fig:speckle}.\\

\noindent{\bf Measurement of $\VecT_{\text{exp}}$}\\
Here we describe how $\VecT_{\text{exp}}$ is measured. The complex element of $\VecT_{\text{exp,mn}}$ in the $n^{\text{th}}$ column and $m^{\text{th}}$ row is found by illuminating the MPLC with the $n^{\text{th}}$ input mode (i.e., $\text{HG}_n$) and holographically measuring the output field, $\Vecc_n$, which is normalised such that the total input power is 1. We project $\Vecc_n$ onto the spot basis with the $m^{\text{th}}$ output spot field defined by $\Vecs_n$, which yields $\VecT_{\text{exp,mn}} = \Vecs_m\cdot\Vecc_n$.

To compare the experimental and target TMs we calculate the overlap between them using:
\begin{equation}
    o = \frac{\left|\VecT_{\text{exp}}:\VecT_{\text{0}}\right|^2}{\left[\VecT_{\text{exp}}:\VecT_{\text{exp}}\right]\left[\VecT_{\text{0}}:\VecT_{\text{0}}\right]},
\end{equation}
where $:$ denotes a scalar (or Frobenius) product, defined for two complex matrices $\VecA, \VecB$ as $\VecA:\VecB = \sum_m\sum_n A_{mn}^*B_{mn}$; here $^*$ denotes a complex conjugate.

\onecolumngrid
\vspace{0.5cm}
\twocolumngrid
 
%

\section{Acknowledgements}
 DBP acknowledges financial support from the European Research Council (ERC): ERC Starting grant {\it PhotUntangle}, no.~804626; and ERC Consolidator grant {\it ModeMixer}, no.~101170907. DBP and UGB thank the Engineering and Physical Sciences Research Council (EPSRC) for financial support: {\it Photon management in complex dynamic scattering media}, EP/Z535928/1; and {\it A-Meta}, EP/W003341/1. DBP also acknowledges support from an EPSRC Hub grant: {\it MetaHub}, UKRI1255. UGB thanks Faidon Kyriakou for photography support.

\newpage

\section{Contributions}
 DBP conceived the idea for the project and developed it with UGB. UGB built the laser writing platform and the MPLC testing system, and wrote all experimental control software. MB provided support on the design of the laser writing platform and advised on the laser writing parameter optimisation. UGB designed all MPLCs, with support from DBP. UGB fabricated and tested the MPLCs and analysed the results. DBP and UGB wrote the paper, with editorial input from MB. DBP obtained the funding and supervised the project.

\newpage

\onecolumngrid
\newpage
\appendix
\clearpage

\renewcommand{\thefigure}{\arabic{figure}}
\renewcommand{\figurename}{Supplementary Figure} 
\renewcommand{\tablename}{Supplementary Table}

\setcounter{figure}{0} 
\setcounter{equation}{0}
\setcounter{table}{0}
\setcounter{page}{1}

\section{\large Supplementary information for}
\section{\large Miniaturised transmissive multi-plane light converters via laser-written\\ geometric phase holograms cascaded in glass}
\vspace{40pt}

\section*{\S1. MPLC characterisation setup}
\begin{figure*}[h]
\includegraphics{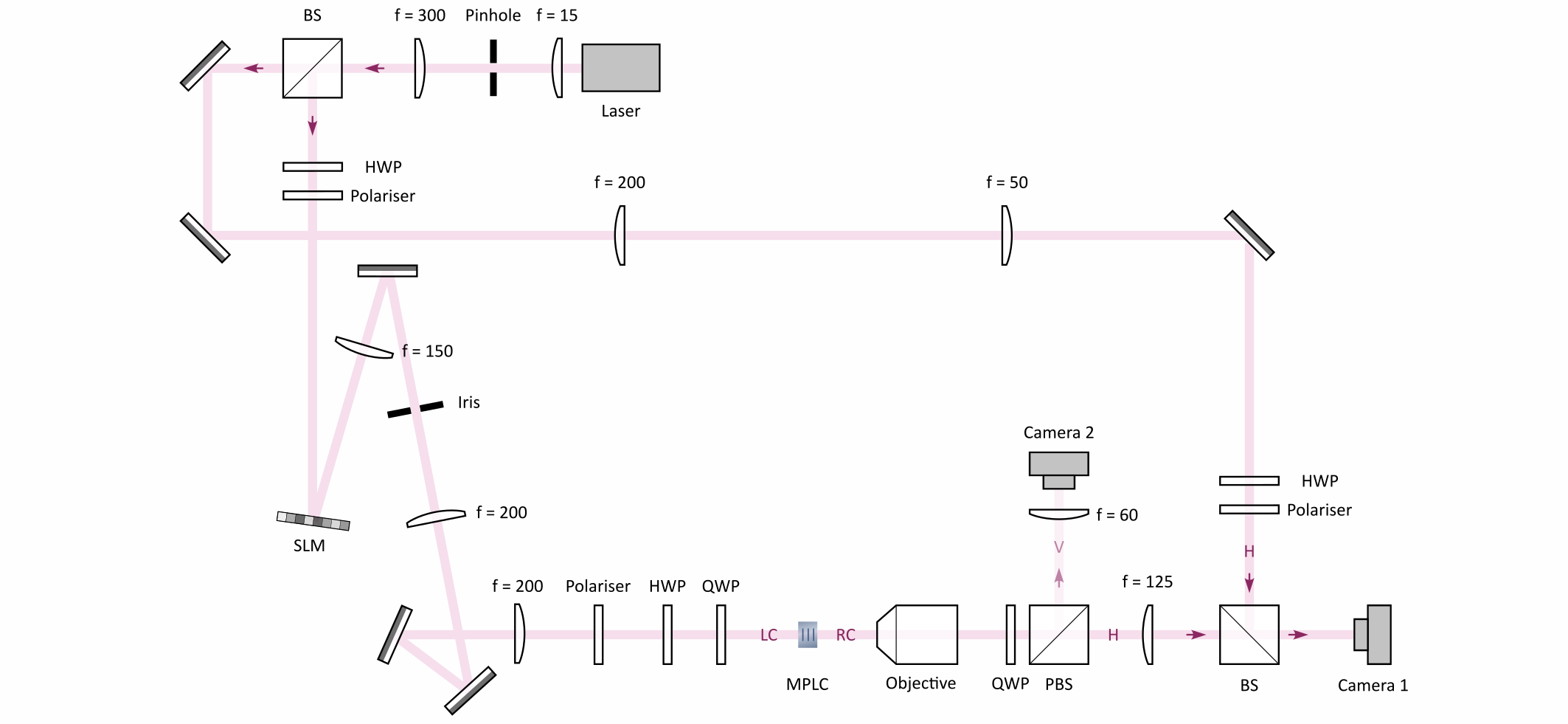}
     \caption{{\bf Optical setup for analysing MPLC performance.} A 633\,nm laser beam is expanded and spatially filtered. A beam splitter (BS) splits the light into signal and reference arms. Energy control is enabled on the signal arm with a rotatable half-wave plate (HWP) and a polariser. The light is then directed onto a liquid crystal spatial light modulator (SLM), where it is shaped to create the input modes. The iris blocks the zero-order light (not illustrated) coming from the SLM. The input modes are re-imaged onto the first plane of the MPLC using a 4f system, and polarisation is set to left-circular (LC) using a polariser, a HWP, and a quarter-wave plate (QWP). The light then interacts with the MPLC. Shown here is a 3-plane MPLC, which means that the outgoing polarisation will be right-circular (RC). Using a QWP we convert the outgoing polarisation back into linear, which we then pass through a polarising beam splitter (PBS). The correct polarisation is directed towards camera 1, which is set up with the objective lens to image the output plane of the MPLC. The reference beam is also directed onto camera 1 for complex field measurements. Camera 2 is set up to image the last plane of the MPLC in order to aid alignment.}
     \label{fig:figSup_readoutSetup}
\end{figure*}

\newpage

\section*{\S2. Cross-talk matrices}
\begin{figure*}[h]
\includegraphics{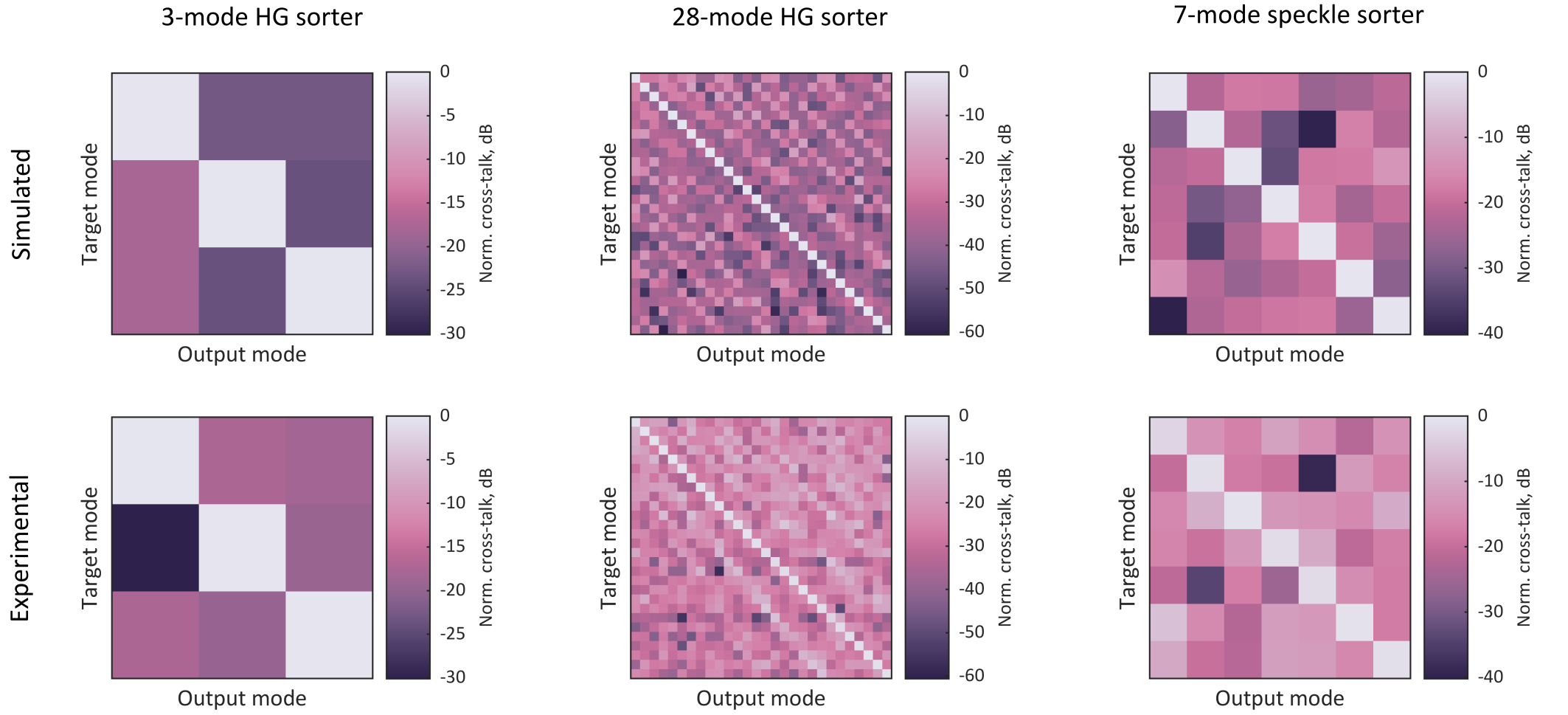}
     \caption{{\bf Simulated and experimentally measured mode sorter cross-talk matrices presented on a log scale.} }
     \label{fig:figSup_logCrosstalk}
\end{figure*}

\section*{\S3. Mode sorting efficiency}
\begin{figure*}[h]
\includegraphics{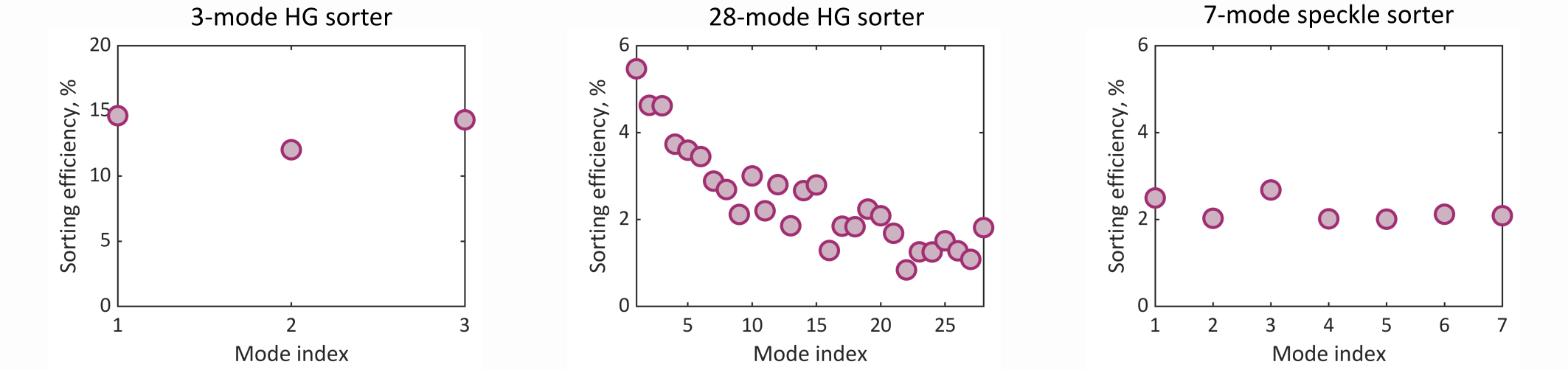}
     \caption{{\bf Overall experimental efficiency of mode sorters.} Here we particularly note that the efficiency of the $28$-mode HG sorter varies as a function of mode index -- with higher order modes tending towards a lower efficiency (even after input intensity normalisation). This is likely due to the fact that higher order HG modes contain higher spatial frequency field components. Since these field components propagate at steeper angles relative to the optical axis of the MPLC, they are more likely to be scattered upon traversing the physically engineered phase masks of finite thickness. We expect the mode sorting efficiency could be balanced by using a more physically accurate model in the inverse design algorithm, and also incorporating an efficiency equalisation term into the design optimisation cost function.}
     \label{fig:figSup_sortingEfficiency}
\end{figure*}

\end{document}